\documentclass[usenatbib,usegraphicx,times]{mn2e}
\include{epsf}

\def\spose#1{\hbox to 0pt{#1\hss}}
\def\simlt{\mathrel{\spose{\lower 3pt\hbox{$\mathchar"218$}}
     \raise 2.0pt\hbox{$\mathchar"13C$}}}
\def\simgt{\mathrel{\spose{\lower 3pt\hbox{$\mathchar"218$}}
          \raise 2.0pt\hbox{$\mathchar"13E$}}}
\def\HI{H{\small\sc{I}~}}

\title[Cosmological simulations of the high-z radio universe]
{Cosmological simulations of the high-redshift radio universe}
\author[D.~Kawata, B.K.~Gibson \& R.A.~Windhorst]
 {D.~Kawata$^{1}$\thanks{E-mail: dkawata@astro.swin.edu.au},
B.K.~Gibson$^{1}$
and R.A.~Windhorst$^{2}$
\\
$^{1}$Centre for Astrophysics \& Supercomputing, 
Swinburne University, Hawthorn VIC 3122, Australia
\\
$^{2}$Department of Physics \& Astronomy, Arizona State University, P.O.~Box
871504, Tempe, AZ 85287-1504, USA
}
\date{Accepted .
      Received ;
      in original form }

\pagerange{\pageref{firstpage}--\pageref{lastpage}}
\pubyear{2004}

\begin{document}

\maketitle

\label{firstpage}

\begin{abstract}
Using self-consistent cosmological simulations of disc galaxy formation, we
analyse the 1.4~GHz radio flux from high-redshift progenitors of present-day
normal spirals within the context of present-day and planned next-generation
observational facilities.  We demonstrate that while current radio facilities
such as the \it Very Large Array \rm (VLA) are unlikely to trace these
progenitors beyond redshifts $z\simgt0.2$, future facilities such as the {\it
Square Kilometer Array} (SKA) will readily probe their characteristics to
redshifts $z\simlt2$, and are likely to provide detections beyond
$z\approx3$.  We also demonstrate that 
the progenitors of present-day cD galaxies
can emit in excess of 10~${\rm \mu Jy}$ of flux at redshifts
$z\simgt1$, and may be a non-negligible contributor to the micro-Jansky
source counts derived from current deep VLA cm-wave surveys.
\end{abstract}


\section{Introduction}
\label{intro-sec}

Astronomy is on the brink of an instrumental revolution. 
The next generation of 
observational facilities will afford in most cases, in excess of 1$-$2 orders
of magnitude improvement in sensitivity and/or resolution across the full
electromagnetic spectrum - from the X-ray ({\it X-Ray Evolving Universe 
Spectrometer}), to the optical ({\it Extremely Large Telescopes}), to
the infrared ({\it James Webb Space Telescope}), to the millimeter
({\it Atacama Large Millimeter Array}), and into the 
radio regime ({\it Square Kilometer Array} - SKA, hereafter).
The SKA\footnote{http://www.skatelescope.org/} 
is expected to deliver sub-$\mu$Jy sensitivity at 1.4~GHz -
$\sim$50$-$100 times that capable with the existing premier
radio facility, the {\it Very Large Array} (VLA) which will be upgraded
to the {\it Expanded Very Large Array} (E-VLA) and whose
sensitivity will increase by factor five before the SKA will be built.

At 1.4~GHz, radio counts are dominated by active galactic nuclei (AGN) at
fluxes $\simgt$ a few mJy, 
while the fainter ($\sim$10~$\mu$Jy) counts are dominated
by starburst galaxies \citep{wmokk85,wfpl93,hwce00}.  
At the anticipated SKA sensitivity limit ($\sim$0.1~$\mu$Jy), little is known 
of the sources which may dominate the radio universe.
Based on an empirical model, \citet{hwce00} predict that normal disc 
galaxies near $z\approx0.5-1$ will dominate the sub-$\mu$Jy counts.
As the radio continuum flux is known to be roughly proportional to 
the star formation rate (SFR) for non-AGN systems \citep{jc92}, one
might anticipate that the SKA would probe the progenitors to these normal disc
galaxies directly, should this prediction be valid.

To test the above hypothesis, we have constructed cosmological simulations of
disc galaxy formation which include a self-consistent treatment of 
temporal evolution of the radio flux.  Our simulations were carried out
using the galactic chemodynamics code {\tt GCD+} \citep{kg03a,kg03b}.
{\tt GCD+} is a three-dimensional tree $N$-body/smoothed particle
hydrodynamics (SPH) code which incorporates self-gravity, hydrodynamics, 
radiative cooling, star formation, supernovae (SNe) feedback, and metal
enrichment. {\tt GCD+} takes account of
the chemical enrichment by both Type~II (SNe~II) and Type~Ia (SNe~Ia)
SNe, mass-loss from intermediate mass stars, and follows the chemical
enrichment history of both the stellar and gas components of the system.
The radio flux is calculated based upon an
empirical law in which the radio flux is linked with the event rate of SNe~II
as described in Section~\ref{arc-sec} \citep{bsg02,ls99}.

The details of {\tt GCD+} relevant to the specific radio source count
simulations described here are presented in Section~\ref{meth-sec}.
In Section \ref{rcev-sec}, we show the expected flux from 
progenitors of simulated disc galaxies and compare the result with both
the current observational data and the anticipated SKA sensitivity limit.
Section~\ref{rcclp-sec} described what we consider to be
plausible candidates for some fraction of the faint radio sources detected in
current faint cm-wavelength radio surveys.  Our results are discussed fully 
in Section~\ref{dc-sec}.

\begin{figure*}
\epsfxsize=160mm  
\epsfbox{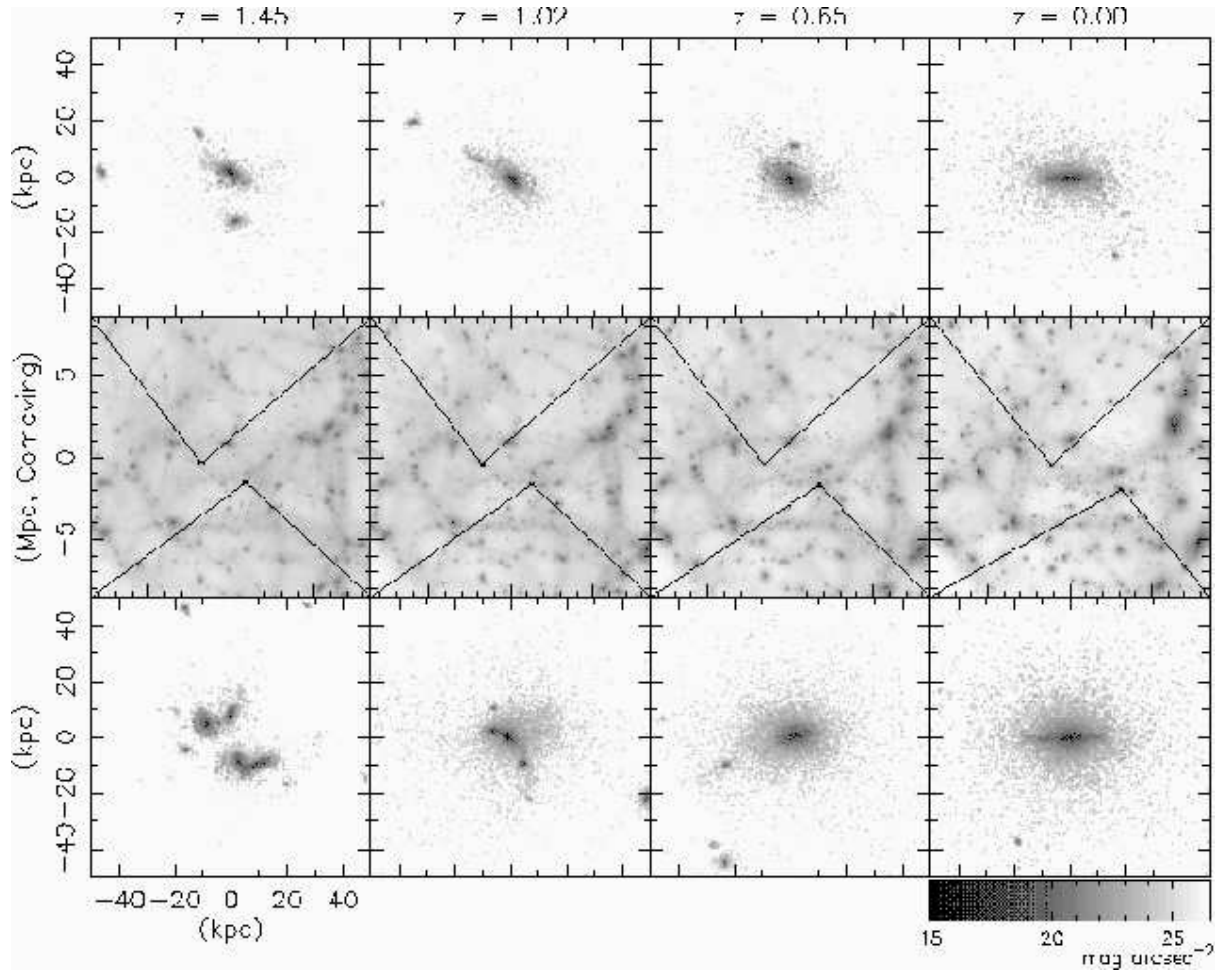}
\caption{Dark matter density map of a portion of 
the simulation volume of a comoving 20$h^{-1}$ Mpc diameter
sphere ({\it middle panels}), 
and predicted $J$-band image (physical scale) of
galaxies D1 ({\it upper panels}) and D2 ({\it lower panels}),
over the redshift range $z$=1.45 to $z$=0. The AB-magnitude system has been
employed.}
\label{evol-fig}
\end{figure*}

\section{Method}
\label{meth-sec}

\subsection{The Galactic Chemodynamical Evolution Code: GCD+}
\label{gcd-sec}

 Our simulations were carried out using {\tt GCD+}, our original
galactic chemodynamical evolution code.
Details of the code are presented in \citet{dk99}
and \citet{kg03a}.
The code is essentially based on {\tt TreeSPH} \citep{hk89,kwh96},
which combines the tree algorithm \citep{bh86}
for the computation of the gravitational forces with the SPH
\citep[][]{l77,gm77}
approach to numerical hydrodynamics.
The dynamics of the dark matter
and stars is calculated by the $N$-body scheme, and the
gas component is modeled using SPH.
It is fully Lagrangian, three-dimensional, and highly adaptive in space
and time owing to individual smoothing lengths and
individual time steps. Moreover, it includes self-consistently
almost all the important physical processes
in galaxy formation, such as self-gravity, hydrodynamics,
radiative cooling, star formation, SNe feedback and
metal enrichment. 

Radiative cooling which depends on the metallicity of the gas
\citep[derived with MAPPINGSIII:][]{sd93} is taken into account. 
The cooling rate for a gas with solar metallicity
is larger than that for gas of primordial composition
by more than an order of magnitude.
Thus, the cooling by metals should not be ignored
in numerical simulations of galaxy formation
\citep{kh98,kpj00}.

 We stress that {\tt GCD+} takes account of energy feedback and
metal enrichment by both SNe II and SNe Ia,
and metal enrichment from intermediate mass stars 
\citep[see][for details]{kg03a}. 
 The code calculates the event rates of SNe II and SNe Ia,
and the yields of SNe II, SNe Ia and intermediate mass stars
for each star particle at every time step,
considering the \citet{es55} initial mass function (mass range of 
0.1$\sim$60 ${\rm M_{\rm \sun}}$) and metallicity dependent stellar lifetimes 
\citep{tk97,ka97}.
We assume that each massive star ($\geq8\ {\rm M_{\rm \sun}}$)
explodes as a Type II supernova.
The SNe Ia rates are calculated using the model 
proposed by \citet{ktn00}.
The yields of SNe II, SNe Ia and intermediate mass stars
are taken from \citet{ww95}, \citet{ibn99} and \citet{vdhg97},
respectively.
The simulation follows the evolution of the abundances
of several chemical elements ($^1$H, $^4$He,$^{12}$C, $^{14}$N, $^{16}$O,
$^{20}$Ne, $^{24}$Mg, $^{28}$Si, $^{56}$Fe, and Z,
where Z means the total metallicity).
{\tt GCD+} makes it possible for us to follow the galaxy formation
processes, taking into account both chemical and dynamical evolution
effects self-consistently, and analyse the observational properties,
such as optical and radio emission, from simulation results. 

\subsection{Cosmological Simulations}
\label{cs-sec}

Using {\tt GCD+}, we carried out a series of high-resolution
simulations within an adopted standard $\Lambda$-dominated cold 
dark matter ($\Lambda$CDM) cosmology,
$\Omega_0$=0.3, $\Lambda_0$=0.7, 
$\Omega_{\rm b}$=0.019$h^{-2}$, $h$=0.7 and $\sigma_8$=0.9
\footnote{These
cosmological parameters are close to 
the most favoured values after
the {\it Wilkinson Microwave Anisotropy Probe} \citep{svp03}.
The differences are no larger than 15 \%.}.
We used a multi-resolution technique in order to maximise the spatial
resolution within the regions where the disc progenitors form and evolve.
Tidal forces from neighbouring
large-scale structure are included self-consistently 
\citep[see also][]{kg03b}. 
The initial conditions for the simulations
were constructed using {\tt GRAFIC2} \citep{eb01}.
All our simulations use isolated boundary conditions.
Initially, we perform a low-resolution $N$-body simulation of a comoving 
20~$h^{-1}$ Mpc diameter sphere which consists of 137376 particles
with the mean separation of 20~$h^{-1}$/64 Mpc. 
The mass of each particle in our low-resolution simulation
was $3.63\times 10^{9} {\rm M_{\sun}}$, and a fixed softening 
of 9~kpc (in physical scale) was applied.

At redshift $z=0$, we selected a 10~Mpc diameter spherical region
which contains several galaxy-sized dark matter halos.
We traced the particles which fall into the selected region
back to the initial conditions at $z=43.5$, and identify the volume 
which consists of those particles. Within this arbitrarily shaped volume,
we replace the low-resolution particles with particles a factor of 64 times
less massive.
The initial density and velocities for the less massive 
particles are self-consistently calculated by {\tt GRAFIC2}, taking into
account the density fields of a lower-resolution region.
Finally, we re-run the simulation of the entire volume 
(20~$h^{-1}$ Mpc sphere), including gas dynamics, radiative cooling,
and star formation. The gas component is included only within the
high-resolution region. The surrounding low-resolution region
contributes to the high-resolution region only through gravity.
The mass and softening length of individual gas (dark matter) particles 
in the high-resolution region are $7.33\times10^6$ 
($4.94\times10^7$) ${\rm M}_\odot$ and 1.14 (2.15) kpc, respectively.

Using the friends-of-friends (FOF) technique, 
we next identified six stellar systems
which consisted of more than 2000 star particles.
Two of these systems possess disc-like morphologies and internal kinematics
consistent with being supported primarily by rotation (and have $>$10000
star particles each).
Table~\ref{gal-tab} summarises the basic properties
of these disc galaxies which we refer to as galaxies D1 and D2 hereafter.
In Table~\ref{gal-tab}, the virial mass $M_{\rm vir}$ is defined as the mass 
within the virial radius $r_{\rm vir}$, where the latter is defined to
be the radius of 
the sphere containing a mean density of
$178 \Omega_0^{0.45}$ times the critical value, 
$\rho_{\rm crit}=3 {\rm H_0} /8\pi G$, following \citet{enf98}.
The rotation velocity $V_{\rm rot}$ was measured directly from the 
rotation curves of the respective gas discs.
Galaxy D1 has a similar rotation velocity to that of
the Milky Way, with a $K$-band luminosity approximately twice that of an
L$^\ast$ late-type spiral \citep[$M_{K*}=-22.98\pm0.06$ mag:][]{kpf01}. 
Because the stellar masses associated with our simulated galaxies are likely
to be overestimated, due to the ubiquitous over-cooling problem
(White \& Frenk 1991), we consider galaxy D1 to be a fair representation
of a typical disc galaxy.  Galaxy D2 is approximately a factor of two more
massive and luminous than D1. 
 Admittedly, a sample of two is not ideal, but we emphasise that this 
is a pilot study for the longer-term goal of better characterising 
the capabilities of the next generation of radio facilities.  
Since progenitors of these two galaxies have similar radio fluxes 
despite their different formation histories, as shown below,
it is re-assuring at least that 
our conclusions are independent of the specifics of the two high-resolution
disc galaxies described here.

\begin{table*}
 \centering
 \begin{minipage}{140mm}
 \caption{Simulated disc galaxy properties.}
 \label{gal-tab}
 \begin{tabular}{@{}lcccccccc}
 Name & ${\rm M_{vir}} $ & ${\rm r_{vir}}$ & $V_{rot}$ & $M_K$ &
 \multicolumn{3}{c}{Number of Particles} \\
   & (${\rm M_{\odot}}$) &  (kpc) & (km s$^{-1}$) & (mag) & 
   Gas & dark matter & Star\\
 D1 & $9.8\times 10^{11} $ & 305 & 220 & $-$23.9 &
  3997  & 17408 & 11819 \\
 D2 & $2.4\times 10^{12} $ & 408 & 230 & $-$24.3 &
 12222 & 41792 & 27636 \\
 \end{tabular}
 \end{minipage}
\end{table*}

Fig.~\ref{evol-fig} shows the evolution of the dark matter distribution
in a central portion of the simulation volume (middle row), 
and the evolution of the stellar component ($J$-band)
in a 100-kpc region centred
on galaxies D1 and D2 (upper and lower rows, respectively). 
Here, the optical luminosities are derived by our population synthesis
code adopting the simple
stellar population of the public spectrum and chemical evolution code,
{\tt KA97} \citep{tk97,ka97}, taking into account the age and metallicity of
star particles \citep[see also][and references therein]{kg03b}. 
We have taken into account $k$-corrections, but have neglected the influence of
dust extinction.
Both galaxies form through conventional hierarchical clustering
by redshift $z$=1; the disc component is built-up via
smooth accretion of gas thereafter.   By the present-day, both 
simulated galaxies have prominent stellar and gaseous discs.

\begin{figure}
\epsfxsize=80mm  
\epsfbox{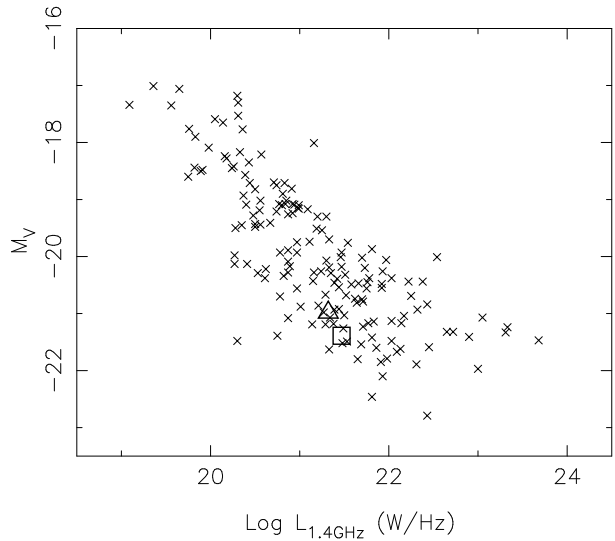}
\caption{
Predicted optical $V$-band magnitude and radio 1.4 GHz luminosities
for galaxies D1 (triangle) and D2 (square).
Crosses show the observational data for nearby galaxies
from \citet{eb03}. 
}
\label{lrmv-fig}
\end{figure}

\subsection{Analysis: Radio Continuum}
\label{arc-sec}

We assume here that a simulated galaxy's radio continuum consists of both
thermal ($L^{\rm T}$) and non-thermal ($L^{\rm NT}$) emission components.
We do not currently include a contribution from AGN.
Following \citet{bsg02}, the non-thermal emission is assumed to 
follow a simple empirical formula related directly to the SNe~II rate
$R_{\rm SNe\ II}$,
\begin{eqnarray}
L^{\rm NT}(\nu) & = &
\left[
0.06\left(\frac{\nu}{1.49 {\rm GHz}}\right)^{-0.5} 
+ 1.38\left(\frac{\nu}{1.49 {\rm GHz}}\right)^{-0.9}
\right] \nonumber \\
& &\times \frac{R_{\rm SNe\ II}}{\rm yr^{-1}}\ 
({\rm erg\ s^{-1}\ Hz^{-1}}),
\label{lnth-eq}
\end{eqnarray}
where $\nu$ is rest-frame frequency (in GHz). Since {\tt GCD+} takes
into account stellar lifetime, $R_{\rm SNe II}$ is different from SFR.
We assume that
the thermal emission follows the simple scaling relation described by
\citet{ls99} \citep[see also][]{cy90},
\begin{equation}
\frac{L^{\rm NT}(\nu)}{L^{TH}(\nu)}  = 10.9 
\left(\frac{\nu}{\rm GHz}\right)^{-0.7}.
\label{lntth-eq}
\end{equation}

While admittedly a simplistic assumption, these empirical laws do 
reproduce successfully the radio flux of observed present-day disc galaxies.
Fig.~\ref{lrmv-fig} shows the predicted optical $V$-band 
and 1.4~GHz fluxes for galaxies D1 (triangle) and D2 (square).
The crosses correspond to
the observational data for nearby galaxies from \citet{eb03}.
As we do not yet have a self-consistent treatment of dust extinction
within {\tt GCD+}, we have de-reddened
the observational data in Fig.~\ref{lrmv-fig} using the extinction
tabulated by \citet{eb03}. 
We assume that the luminosity within an aperture of 50~kpc 
(in physical scale) radius encompasses 
the entire optical and radio-emitting regions.
Fig.~\ref{lrmv-fig} shows that
the predicted optical and radio luminosities for the simulated galaxies
are consistent with those of observed galaxies. 
Encouraged by this, below we assume that this empirical law 
is also valid for high-redshift galaxies.
In the next section, we discuss the temporal evolution of the
predicted radio flux for the
progenitors of our two simulated disc galaxies.

\begin{figure}
\epsfxsize=80mm  
\epsfbox{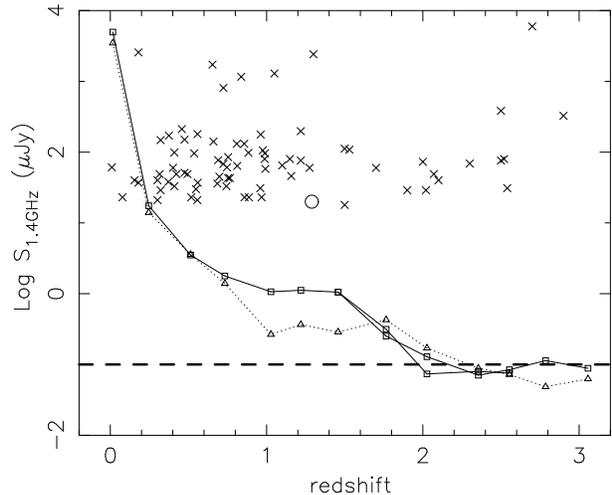}
\caption{
The predicted 1.4~GHz flux in the observed frame from 
progenitors of the simulated 
galaxies D1 (triangles connected by doted line) and D2 
(squares connected by solid line) at different redshifts. 
Crosses correspond to the observational data in \citet{hpwr00}. 
The open circle represents the predicted flux
from a progenitor of a present-day (simulated) cD galaxy in our model.
Thick dashed line indicates the anticipated SKA flux sensitivity limit
(0.1 ${\rm \mu Jy}$) in 12 hours.
}
\label{zaf-fig}
\end{figure}

\section{Results}

\subsection{Radio flux from disc progenitors}
\label{rcev-sec}

Fig.~\ref{zaf-fig} shows the predicted 1.4~GHz flux (observer's frame)
from progenitors
of galaxies D1 and D2 at different redshifts. Here, we measure 
the total flux within an aperture of 50~kpc (in physical scale)
radius and take into account
the $k$-correction.
At each redshift, we use the FOF to find
the stellar systems identified as the progenitors of galaxies D1 and D2.
Only progenitors which consist of more than 2000 star particles are
displayed in Fig.~\ref{zaf-fig}. At $z=1.8$, galaxy D2
has two progenitors, which by redshift $z=1.4$ have merged.
Actually, even by visual inspection, galaxy D2 appears to have more than
two progenitors still identifiable at $z=1.45$, all of which could be 
resolved with the SKA arcsecond spatial 
resolution\footnote{10~kpc at redshift $z=1.45$ corresponds to
 $\sim$1.2 arcseconds in the adopted cosmology.}
\citep[see also][]{cgor97}.
However, as our simple FOF classified them as one system, 
we simply define such system as a single object in the accompanying
figure. 

The aperture radius of $R_{\rm ap}=$50 kpc has been chosen arbitrarily.
This is roughly a radius which encompasses the entire 
stellar system identified as a single stellar system by our FOF
at any redshift which we focus on. To see the aperture effect,
we also measured the fulxes within $R_{\rm ap}=100$ and 10 kpc. 
At $z<1$, the flux within $R_{\rm ap}=100$ kpc agrees with
that within $R_{\rm ap}=50$ kpc within about 10 \%,
while the flux within $R_{\rm ap}=10$ kpc is systematically
(about 15 \% in the worst case) smaller than
that within $R_{\rm ap}=50$ kpc, because the stellar systems are
bigger than 10 kpc (see Fig.\ \ref{evol-fig}).
At $z>1$, since the central galaxy is relatively small, the contribution
from the satellite galaxies are not negligible. 
Consequently, $R_{\rm ap}=100$ kpc leads to 
50 \% larger flux than $R_{\rm ap}=50$ kpc in the worst case.
On the other hand, in the multiple merger like the progenitors of galaxy D2 at 
$z=1.45$ (Fig.\ \ref{evol-fig}), $R_{\rm ap}=10$ kpc sometimes leads to
about 20 \% of flux within $R_{\rm ap}=50$ kpc for the biggest
galaxy. Note that
the radio interferometer, such as the SKA, enables to
resolve them, and they are likely detected as multiple sources,
as mentioned above.

Our quoted SNe~II rates are averages over the most recent
1 Gyr for any system at $z<1$ and 0.05 Gyr for the ones at $z>1$.  
While a somewhat
arbitrary ``binning'', we confirm that our results are not
sensitive to the specific bin-size adopted.  Such a sampling
suppresses the stochastic appearance of relatively rare 
short period ($\sim$Myr) starbursts which are likely caused by
resolution limits.
Thus, the results of Fig.~\ref{zaf-fig} represent a mean trend in
the expected flux from progenitors of galaxies D1 and D2, as
implicitly supported by the smooth nature of the curves in the
accompanying figure.

The crosses in Fig.~\ref{zaf-fig} correspond to the observed VLA
dataset of \citet{hpwr00}.
The flux limit for this dataset is approximately 10~${\rm \mu Jy}$.
Our simulations suggest that the VLA count source data corresponds to
the (recent - $z<0.2$) progenitors
of present-day disc galaxies. This is consistent with the conclusions
drawn by optical follow up for the same sources by
\citet{rlk02}.  At higher-redshifts, these progenitors possess radio fluxes
significantly below the feasible 1.4~GHz flux limit of the VLA.
Their predicted flux reaches $\sim$1~${\rm \mu Jy}$ at redshift 
$z\approx1$,
and decreases to $\sim$0.1~${\rm \mu Jy}$ by redshift $z=2$ (with
little evolution at redshifts $z>2$).  We do not trace the
progenitors beyond redshift $z\simgt3$, as the number of bound
star particles become too few ($<$2000) for the results to remain reliable.
Galaxies D1 and D2 show very similar trends
for the temporal evolution of the observed 1.4~GHz flux.
Based upon our (admittedly limited) sample of two high-resolution simulated
disc galaxies, 
we conclude that the trends shown by the curves in 
Fig.~\ref{zaf-fig} are fairly representative of typical disc systems.
Consequently, our results suggest that the SKA is capable of
detecting the progenitors of normal disc galaxies at redshifts 
$z\simlt3$, provided it meets its design specifications of 
0.1~${\rm \mu Jy}$ \citep[the expected 5 $\sigma$ level achievable after
a 12 hour integration, e.g.][]{hwce00} as shown by the thick dashed line
in Fig.~\ref{zaf-fig}.

\begin{figure}
\epsfxsize=80mm  
\epsfbox{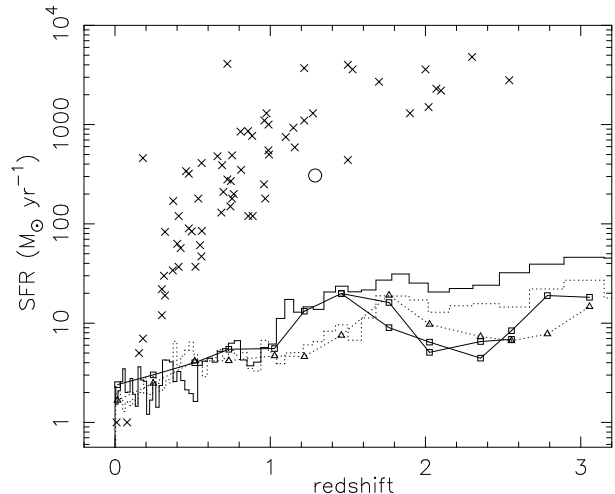}
\caption{
The SFR for progenitors of the simulated galaxies D1 
(triangles connected by doted line) and D2 
(squares connected by solid line) at different redshifts. 
Crosses correspond to the observational data from \citet{hpwr00}. 
Dotted and solid histograms show
the history of the total SFR for galaxies D1 and D2 (see text for details).
The open circle shows the SFR for a progenitor to a present-day 
(simulated) cD galaxy. 
The reason why there are no observational data with low SFR at high
redshift is the observational selection effect due to the limited sensitivity.
}
\label{zsfr-fig}
\end{figure}

Fig.~\ref{zsfr-fig} shows the SFR of the progenitors of galaxies 
D1 and D2 at different redshifts.  Here, the quoted SFR has been derived by
averaging over the same aperture and temporal bin as that described
previously.  The crosses represent the inferred SFR of
the VLA faint sources from \citet{hpwr00},
who estimates the SFR using an empirical relation between SFR 
and radio continuum flux \citep{jc92}. 
Because of the adopted empirical link between radio flux and SNe~II rate
(eq.~[\ref{lnth-eq}]), the latter of which also relates directly to 
the SFR, Figs.~\ref{zaf-fig} and \ref{zsfr-fig} provide somewhat
complementary information.
The reason why there are no observational data with low SFR at high
redshift is the observational selection effect due to the limited
sensitivity, which is the same as the effect which leads to
the well-known envelopes of observational data 
in the absolute radio  luminosity--redshift 
plane \citep[e.g.\ Fig.\ 9 of][]{grh04}.
Fig.\ \ref{zsfr-fig} suggests that the
progenitors of galaxies D1 and D2
have much smaller SFRs than the VLA faint sources at $z>0.2$
(which is simply another way of stating why they have much 
lower radio fluxes).

The dotted and solid histograms of Fig.~\ref{zsfr-fig} show 
the history of the total SFR for galaxies D1 and D2, respectively.
Here, the total SFR means the summation of the SFR of
star particles which fall into the region within a galactocentric 
radius of 50~kpc at redshift $z=0$. For both galaxies,
the total SFR is similar to the SFR of the most massive progenitor at 
redshift $z\simlt1.5$. 
However, at $z\simgt1.5$ the SFR of the most massive
progenitor is significantly lower than the total SFR. 
This difference indicates that a significant number of stars 
are forming in smaller progenitors which later accrete onto the
most massive progenitor.  An obvious conclusion from this cursory 
analysis is that in order to properly estimate the observed flux from
high-redshift objects, 
it is (not surprisingly) important to take into account
the hierarchical clustering history of the systems.
Galaxy D2 is a clear example - at $z>1.5$, we have noted previously
the presence of
two massive progenitors, and the fact that we are able to 
analyse the radio flux from each progenitor separately. 
This demonstrates that numerical simulations can be a powerful tool for
studying the anticipated
radio flux from individual galactic building blocks.

\begin{figure}
\epsfxsize=80mm  
\epsfbox{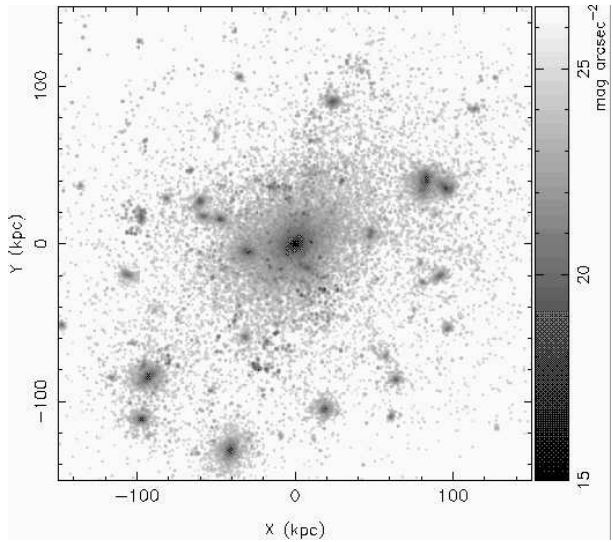}
\caption{
$J$-band (in the observed frame, and using the AB-magnitude system) 
image of a progenitor of a cluster of galaxies.
}
\label{cljab-fig}
\end{figure}

\subsection{Radio flux from cD galaxy progenitors}
\label{rcclp-sec}

In the previous section, we found that the VLA sources observed at $z>0.2$
are unlikely to be progenitors of normal disc galaxies.  The obvious
question then arises as to which population they belong.  We suggest that one
candidate might be the
progenitors to bright central-dominant (cD) cluster galaxies.
Using the same multi-resolution
technique described in Section~\ref{cs-sec}, we carried out a high-resolution
simulation of a cluster of galaxies which was similarly identified first
from a low-resolution cosmological simulation
of a comoving 40~$h^{-1}$ Mpc diameter sphere, and whose virial mass
at redshift $z=0$ was $1.2\times10^{14}$~M$_{\rm \sun}$. 
The mass and softening length of individual gas (dark matter) particles 
in the high-resolution region were $5.86\times10^7$ 
($3.95\times10^8$) ${\rm M}_\odot$ and 2.27 (4.29) kpc, respectively.

The $J$-band image of the simulated cluster redshift $z=1.3$ is shown
in Fig.~\ref{cljab-fig}.
The basic properties of the cluster at this redshift
are summarised in Table~\ref{cl-tab}.
The cluster has a massive central galaxy - 
a progenitor of what will become a present-day 
cD galaxy - and a number of associated member galaxies. 
The 1.4~GHz flux and SFR within the centre 50~kpc (in physical scale)
of the cluster are 
denoted as open circles in Figs.~\ref{zaf-fig} and \ref{zsfr-fig}.
As seen in Fig.~\ref{cljab-fig}, the flux within this region
is dominated by the flux from the progenitor of the cD galaxy itself.
Figs.~\ref{zaf-fig} and \ref{zsfr-fig} demonstrate that
this cD galaxy progenitor has a sufficiently high SFR, and consequently
a sufficiently high 1.4~GHz flux, to reach the present-day VLA
detection limit.
Therefore, we suggest that the progenitors of present-day cD galaxies may
be lurking within the faint VLA source counts.
Fig.~\ref{zsfr-fig} shows that the SFR of this cD progenitor
is $\sim$300~M$_{\rm \sun}$~yr$^{-1}$. Since SFRs of this magnitude are 
likely to be accompanied by associated dust, the intrinsic ultraviolet
and optical light would suffer from consequent extinction, re-appearing
as sub-millimeter emission.
Thus, the VLA faint sources which are also detected
by the {\it Sub-millimeter Common User Bolometric Array}
\citep[e.g.][]{cbis03} may be counterparts of our simulated cD galaxy.
This result is consistent with the previous studies 
\citep[e.g.][]{gbf98,kn02,nch04}
which demonstrate that high redshift starburst galaxies, 
such as Lyman break galaxies, likely evolve into large galaxies 
in clusters and groups at z=0.

\begin{table*}
 \centering
 \begin{minipage}{140mm}
 \caption{Properties of a cluster progenitor at $z=1.3$.}
 \label{cl-tab}
 \begin{tabular}{@{}ccccccc}
  ${\rm M_{vir}} $ & ${\rm r_{vir}}$ & 
 $\sigma$\footnote{Velocity dispersion around $R=5$ kpc.} & 
 $M_K$\footnote{$K$-band absolute magnitude in the rest frame.} &
 \multicolumn{3}{c}{Number of Particles} \\
   (${\rm M_{\odot}}$) & (kpc) & (km s$^{-1}$) & (mag) &
 Gas & dark matter & Star\\
 $6.3\times 10^{11} $ & 544 & 185 & $-$28.2 &
 59365 & 140947 & 70793 \\
 \end{tabular}
 \end{minipage}
\end{table*}

\section{Discussion and Conclusions}
\label{dc-sec}

Combining our cosmological simulations with an empirical law for
radio continuum flux emission has enabled us to trace the temporal
and spatial evolution of radio flux from star-forming disc and central
dominant cluster galaxies.
We have found that current radio telescopes, including the VLA, are
capable of tracing the 1.4~GHz flux of the progenitors of today's
disc galaxies only locally ($z\simlt0.2$).
Progenitors of normal disc galaxies have a 1.4~GHz flux 
$\simlt$10~${\rm \mu Jy}$ at $z>0.2$, with their predicted flux
reaching $\sim$0.1~${\rm \mu Jy}$ at $z=2-3$.
Since the flux limit of the SKA at 1.4~GHz is expected to be 
$\sim$0.1~${\rm \mu Jy}$ in 12 hours, it should be capable of 
detecting disc galaxy progenitors at redshifts $z\simlt3$.
In addition, from a cluster simulation, we also found that 
progenitors of cD galaxies can possess a 1.4~GHz flux in excess of
10~${\rm \mu Jy}$ 
at $z>1$, and may be a candidate for some fraction of 
the faint VLA sources at $z>1$.

It is worth reminding the reader that hierarchical clustering often 
leads to the presence of more than one significant
progenitor (or building block)
for any one disc galaxy at high redshift.  Our simulated galaxy D2
provides a typical example of such a system.
We predict that the radio flux from these individual building blocks 
will be observable by the SKA.  At the sensitivity limit probed by the SKA,
these building blocks of normal galaxies may dominate the radio
source counts, as they do in the optical faint blue galaxy counts
\citep{dfc98}.
Observations of
such extremely faint radio continuum sources may therefore 
provide a unique window into the accretion history of galaxies.
In addition, the SKA may be able to provide 
redshift information for the sources through \HI line observations
\citep[e.g.][]{rb96}. Then, the SKA would be capable of providing
the luminosity function of radio sources at different redshifts
\citep[e.g.][]{cbfc04}, which would give strong constraints 
on disc galaxy formation scenarios.
To interpret such observations, theoretical models
are important \citep[e.g.][]{tkk01,nytn02}; our pilot study
shows that self-consistent cosmological simulations will
provide useful theoretical support for such analyses.
Encouraged by our preliminary results, 
we are undertaking a series of larger volume 
simulations (anticipated completion in 2005), in order to improve the 
statistics associated with the program described here.
 
One note of caution is that the current numerical simulations
suffer from the well-documented 
over-cooling problem \citep{wf91}, which results in an unavoidably high
SFR in the high-redshift building blocks, and a consequent underestimation
of the SFR at later epochs.  In tandem, this leads to disc systems which
suffer from having overly massive stellar halos and bulges, at the
expense of a prominent thin disc.
Improved modeling of SNe feedback is believed to be crucial
in overcoming this long-standing
problem, and it has been suggested that stronger SNe 
feedback than that adopted here may lead to end-products which 
more closely resemble observed disc galaxies today 
\citep{tc01,sgp03,rysh04,bkgf04}. 
Having said that, even models which do incorporate strong feedback only
result in SFR histories a factor of two different from the canonical
models presented here
\citep[e.g.\ Fig.\ 2 of][]{bkgf04}.  
Hence, this particular problem unlikely impact
our predicted radio flux by more than a factor of two.
Nevertheless, more realistic numerical models of galaxy
formation remain the constant target for our group (and others), and
are a necessity for contributing to the design and anticipated 
performance of next-generation facilities such as the SKA.

\section*{Acknowledgements}

We are grateful to Nobuo Arimoto and Tadayuki Kodama
for kindly providing the tables of their SSPs data.
We acknowledge the Yukawa Institute Computer Facility,
the Astronomical Data Analysis Center of the National Astronomical
Observatory, Japan (project ID: rmn12a), 
the Australian and Victorian Partnership for Advanced
Computing, where the numerical computations for this paper were performed.
BKG and DK acknowledge the financial support of the Australian Research
Council. RAW was supported by NASA/JWST Grant NAG5-12460.

\label{lastpage}


\begin{thebibliography}{999}
\bibitem[Barnes \& Hut (1986)]{bh86}
 Barnes J.E., Hut P., 1986, Nature, 324, 446
\bibitem[Bell (2003)]{eb03}
 Bell E. 2003, ApJ, 586, 794
\bibitem[Bertschinger (2001)]{eb01}
 Bertschinger E., 2002, ApJS, 137, 1
\bibitem[Braun (1996)]{rb96}
 Braun R., 1996, in Raimond E., Genee R., eds,
 The Westerbork Observatory, Continuing Adventure in Radio Astronomy.
 Kluwer, Dordrecht, p. 167
\bibitem[Bressan, Silva \& Granato (2002)]{bsg02}
 Bressan A., Silva L., Granato G.L., 2002, A\&A, 392, 377
\bibitem[Brook et al.\ (2004)]{bkgf04}
 Brook C.B., Kawata D., Gibson B.K., Flynn C., 2004, MNRAS,
 349, 52
\bibitem[Chapman et al.\ (2003)]{cbis03}
 Chapman S.C., Blain A.W., Ivison R.J., Smail I.R., 2003, Nature,
 433, 695
\bibitem[Colley et al.\  (1997)]{cgor97}
 Colley W.N., Gnedin O.Y., Ostriker J.P., Rhoads J.E., 1997, ApJ,
 488, 579
\bibitem[Condon (1992)]{jc92}
 Condon J.J., 1992, ARA\&A, 30, 575
\bibitem[Condon \& Yin (1990)]{cy90}
 Condon J.J., Yin Q.F., 1990,  ApJ, 357, 97
\bibitem[Cowie et al.\ (2004)]{cbfc04}
 Cowie L.L., Barger A.J., Fomalont F.B., Capak P., 2004, ApJL, 603, 69
\bibitem[Driver et al.\ (1998)]{dfc98}
 Driver S.P., Fernandez-Soto A., Couch W.J., Odewahn S.C., Windhorst R.A.,
 Phillips S., Lanzetta K., Yahil A., 1998, ApJL, 496, 93
\bibitem[Eke, Navarro \& Frenk (1998)]{enf98}
 Eke V.R., Navarro J.F., Frenk C.S., 1998, ApJ, 503, 569
\bibitem[Governato et al.\ (1998)]{gbf98}
 Governato F., Baugh C.M., Frenk C.S., Cole S., Lacey C.G., Quinn T., 
 Stadel J., 1998, Nature, 392, 359
\bibitem[Gilbert et al.\ (2004)]{grh04}
 Gilbert G.M., Riley J.M., Hardcastle M.J., Crston J.H., Pooley G.G.,
 Alexander P., 2004, MNRAS, 351, 845
\bibitem[Gingold \& Monaghan (1977)]{gm77}
 Gingold R.A., Monaghan J.J., 1977, MNRAS, 181, 375
\bibitem[Haarsma et al.\ (2000)]{hpwr00}
 Haarsma D.B., Partridge R.B., Windhorst R.A., Richards E.A.,
 2000, ApJ, 544, 641
\bibitem[Hernquist \& Katz (1989)]{hk89}
 Hernquist L., Katz N., 1989, ApJS, 70, 419
\bibitem[Hopkins et al. (2000)]{hwce00}
 Hopkins A., Windhorst R., Cram L., Ekers R., 2000, ExA, 10, 419
\bibitem[Iwamoto et al.\ (1999)]{ibn99}
 Iwamoto K., Brachwitz F., Nomoto K., Kishimoto N.,
 Umeda H., Hix W.R., Thielemann F.-K., 1999, ApJS, 125, 439
\bibitem[Katz, Weinberg \& Hernquist (1996)]{kwh96}
 Katz N., Weinberg D.H., Hernquist L., 1996, ApJS, 105, 19
\bibitem[Kawata (1999)]{dk99} 
 Kawata D., 1999, PASJ, 51, 931
\bibitem[Kawata \& Gibson (2003a)]{kg03a} 
 Kawata D., Gibson B.K., MNRAS, 2003a, 340, 908
\bibitem[Kawata \& Gibson (2003b)]{kg03b} 
 Kawata D., Gibson B.K., MNRAS, 2003b, 346, 135
\bibitem[K\"aellander \& Hultman (1998)]{kh98}
 K\"aellander D., Hultman J., 1998, A\&A, 333, 399
\bibitem[Kay et al. (2000)]{kpj00}
 Kay S.T., Pearce F.R., Jenkins A., Frenk C.S.,
\bibitem[Kobayashi, Tsujimoto \& Nomoto (2000)]{ktn00}
 Kobayashi C., Tsujimoto T., Nomoto K., 2000, ApJ, 539, 26
\bibitem[Kochanek et al.\ (2001)]{kpf01}
 Kochanek C.S., Pahre M.A., Falco E.E., Huchra J.P., Mader J., 
 Jarrett T.H., Chester T., Cutri R., Schneider S.E., 2001, ApJ, 560, 566
\bibitem[Kodama (1997)]{tk97} 
 Kodama T., 1997, Ph.D.\ thesis, University of Tokyo
\bibitem[Kodama \& Arimoto (1997)]{ka97} 
 Kodama T., Arimoto N., 1997, A\&A, 320, 41
\bibitem[Lucy (1977)]{l77}
 Lucy L.B., 1977, AJ, 82, 1013
\bibitem[Nagashima et al.\ (2002)]{nytn02}
 Nagashima M., Yoshii Y., Totani T., Gouda N., 2002, ApJ, 578, 675
\bibitem[Nagamine (2002)]{kn02}
 Nagamine K., 2002, ApJ, 564, 73
\bibitem[Nagamine et al.\ (2004)]{nch04}
 Nagamine K., Cen R., Hernquist L., Ostriker J.P., Springel V.,
 2004, submitted to ApJ (astro-ph/0406032)
\bibitem[Robertson et al.\ (2004)]{rysh04}
 Robertson B., Yoshida N., Springel V., Hernquist L., 2004, ApJ,
  606, 32
\bibitem[Roche, Lowenthal \& Koo (2002)]{rlk02}
 Roche N.D., Lowenthal J.D., Koo D.C., 2002, MNRAS, 330, 307
\bibitem[Salpeter (1955)]{es55}
 Salpeter E.E., 1955, ApJ, 121, 161
\bibitem[Silva (1999)]{ls99}
 Silva L., 1999, PhD Thesis, SISSA/ISAS, Trieste, Italy
\bibitem[Sommer-Larsen, G\"otz \& Portinari (2003)]{sgp03}
 Sommer-Larsen J., G\"otz M., Portinari L., 2003, 596, 47
\bibitem[Spergel et al.\ (2003)]{svp03}
 Spergel D.N., Verde L., Peiris H.V., Komatsu E., Nolta M.R.\ et al.,
 2003, ApJS, 148.175
\bibitem[Sutherland \& Dopita (1993)]{sd93}
 Sutherland R.S., Dopita M.A., 1993, ApJS, 88, 253
\bibitem[Takeuchi et al.\ (2001)]{tkk01}
 Takeuchi T.T., Kawabe R., Kohno K., Nakanishi K., Ishii T.K., Hirashita H.,
 Yoshikawa K., 2001, PASP, 113, 586 
\bibitem[Thacker \& Couchman (2001)]{tc01}
 Thacker R.J., Couchman H.M.P., ApJL, 555, 17
\bibitem[van den Hoek \& Groenewegen (1997)]{vdhg97}
 van den Hoek L.B., Groenewegen M.A.T., 1997, A\&AS, 123, 305
\bibitem[White \& Frenk (1991)]{wf91}
 White S.D.M., Frenk C.S., 1991, ApJ, 379, 52
\bibitem[Windhorst et al.\ (1993)]{wfpl93}
 Windhorst R.A., Fomalont E.B., Partridge B.R., Lowenthal J.D.,
 1993, ApJ, 405, 498
\bibitem[Windhorst et al.\ (1985)]{wmokk85}
 Windhorst R.A., Miley G.K., Owen F.N., Kron R.G., Koo D.C.,
 1985, ApJ, 289, 494
\bibitem[Woosley \& Weaver (1995)]{ww95}
 Woosley S.E., Weaver T.A., 1995, ApJS, 101, 181
\end{thebibliography}
\end{document}